\documentclass[twocolumn]{article}

\usepackage{abstract}
\usepackage{graphicx}
\usepackage[affil-it]{authblk}
\usepackage{epsfig}
\usepackage[latin1]{inputenc}
\usepackage{color}
\usepackage{times}
\usepackage{natbib}
\usepackage{setspace}
\definecolor{red}{rgb}{1,0.,0.}

\newcommand{\morgana}{{\sc morgana }}
\newcommand{\andrew}{{\sc galacticus }}
\newcommand{\munich}{{\it Munich }}
\newcommand{\oldurham}{{\it Durham} }
\newcommand{\oldurlike}{{\it Durham}-like }
\newcommand{\durham}{{\sc galacticus}-CP }
\newcommand{\moddur}{{\sc galacticus}-IP }

\def\lesssim{\lower.5ex\hbox{$\; \buildrel < \over \sim \;$}}
\def\gtrsim{\lower.5ex\hbox{$\; \buildrel > \over \sim \;$}}
\begin{document}

\title{A Research Note on the Implementation of Star Formation and Stellar Feedback in Semi-Analytic Models} 
\author{Fabio Fontanot}
\affil{HITS-Heidelberger Institut f\"ur Theoretische Studien, Schloss-Wolfsbrunnenweg 35, 69118, Heidelberg, Germany}
\affil{Institut f\"ur Theoretische Physik der Universit\"at Heidelberg, Philosophenweg, 16, 69120, Heidelberg, Germany}
\author{Gabriella De Lucia}
\affil{INAF-Osservatorio Astronomico di Trieste, Via Tiepolo 11, I-34143, Trieste, Italy}
\author{Andrew J. Benson}
\affil{Carnegie Observatories, 813 Santa Barbara Street, Pasadena, California, 91101 USA}
\author{Pierluigi Monaco} 
\affil{Dipartimento di Fisica, sez. Astronomia, Universit\'a di Trieste, via G.B. Tiepolo 11, I-34143, Trieste, Italy}
\author{Michael Boylan-Kolchin}
\affil{Center for Cosmology, Department of Physics and Astronomy, 4129 Reines Hall, University of California, Irvine, CA 92697, USA} 

\twocolumn[
\maketitle

\begin{onecolabstract} 
We study the impact of star formation and stellar feedback
prescriptions on galaxy properties predicted by means of
``stripped-down'' versions of independently developed semi-analytic
models (SAMs). These include cooling, star formation, feedback from
supernovae (SNe) and simplified prescriptions for galaxy merging, but
no chemical evolution, disc instabilities or AGN feedback. We run
these versions on identical samples of dark matter (DM) haloes
extracted from high-resolution $N$-body simulations in order to
perform both statistical analysis and object-by-object comparisons. We
compare our results with previous work based on stripped-down versions
of the same SAMs including only gas cooling, and show that all
feedback models provide coherent modifications in the distribution of
baryons between the various gas phases. In particular, we find that
the predicted hot gas fractions are considerably increased by up to a
factor of three, while the corresponding cold gas fractions are
correspondingly decreased, and a significant amount of mass is ejected
from the DM halo. Nonetheless, we also find relevant differences in
the predicted properties of model galaxies among the three SAMs: these
deviations are more relevant at mass scales comparable to that of our
own Galaxy, and are reduced at larger masses, confirming the varying
impact of stellar feedback at different mass scales. We also check the
effect of enhanced star formation events (i.e. starbursts modes),
defined in connection with galaxy mergers. We find that, in general,
these episodes have a limited impact in the overall star formation
histories of model galaxies, even in massive DM halos where
merger-driven star formation has often been considered very
important.


{\it email:} fabio.fontanot@h-its.org
\end{onecolabstract}
]

\section{Introduction}\label{sec:intro}

In order to understand the complex process of galaxy formation over
the entire cosmic history of the Universe, many different physical
processes need to be taken into account. These processes act on
different scales and their interplay appears critical for an
appropriate description of the chain of events leading to the build up
of present-day galaxy population. However, our comprehension of the
physical processes acting on the baryonic components of these haloes
is still limited.

A number of theoretical methods have been introduced, trying to get a
better understanding of galaxy formation and evolution: among these
different methods semi-analytic models (hereafter SAMs - for a review,
see e.g. \citealt{Baugh06}) of galaxy formation and evolution have
become a widely used tool, thanks to their flexibility and
(relatively) low computational costs. In these models, relevant
physical processes are included by assuming empirically and/or
theoretically motivated prescriptions, coupled through a set of
differential equations that describe the mass and energy flows between
the different galactic components (i.e. halo, bulge and disc) and
baryonic phases (i.e. stars, hot and cold gas).

A number of competing models have been proposed, assuming different
(but equally plausible) descriptions of the relevant
processes. Although the analysis of discrepancies and similarities
between predictions from independently developed SAMs has been the
subject of a number of recent studies (see e.g. \citealt{Fontanot09b,
  Fontanot10, DeLucia11} and discussions therein), the role played by
the overall SAM ``architecture'' (i.e. the more technical details of
the construction of the models, and the interplay of their different
assumptions) has not been analysed in detail.

A first step in this direction was given in \citet[hereafter
  DL10]{DeLucia10}. In this study, we used ``stripped-down'' versions
of three independently developed SAMs including only gas cooling and
galaxy mergers. The models were run on identical sets of merger trees
extracted from $N$-body simulations, so as to remove any systematic
effect due to the adopted description of the dark matter evolution. We
found a reasonable level of agreement between predictions from the
three models, for the physical processes considered. In particular,
the agreement is very good at dark matter haloes (DMHs) scales
comparable to those of our own Milky-Way (MW). In the other hand, for
larger masses, corresponding to those of clusters at $z=0$, we found
significantly different results in the predicted amount of cold gas,
largely due to the different assumptions for the hot gas distribution
inside DMHs, and to different treatments of the ``rapid cooling''
regime.

In DL10, we also compared the different treatment for the dynamical
evolution of substructures and galaxy mergers, based either on fitting
formulae derived from numerical simulations or analytic models
accounting for dynamical friction, tidal stripping and tidal
shocks. We showed that these different assumptions result in
significant differences in the timings of mergers, with important
consequences for the formation and evolution of massive galaxies. 

In this research note, we extend the analysis of DL10 to investigate
the influence of different prescriptions adopted for the physical
mechanisms of star formation and SN feedback in both ``quiescent'' and
``starburst'' modes, avoiding to consider other processes like metal
evolution, AGN feedback and disc instabilities and using simplified
treatment of galaxy mergings. We stress that, given our limited
understanding of the physical processes considered, all the
prescriptions we will discuss in the following are equally plausible,
so our analysis is not aimed at identifying the ``best'' model for
star formation and feedback. Rather, the our aim is to analyse the
influence of different model ingredients (and of the various modelling
that can be adopted for specific processes) on the predicted
properties of galaxies and their redshift evolution. As in DL10, we
run our models on the same sets of merger trees extracted from
numerical simulations. Results from our previous study allow us to
control residual differences due to a different model for gas cooling
and galaxy mergers.

In this work we use stripped-down versions of the \munich model by
\citet{DeLuciaBlaizot07}, the \andrew model of \citet{Benson12} as an
extension of \oldurham model of \citet{Bower06}, and the \morgana
model of \citet{Monaco07}. In all models we implement radiative
cooling of a gas with primordial composition, star formation and SN
feedback in ``quiescent'' and ``starburst'' regimes, including models
of ``super-wind'' ejection from the DM halos and later re-accretions,
and simplified prescriptions for galaxy mergers. We do not modify the
choice of parameter values with respect to the original calibrations:
we do not then expect predictions from these stripped-down versions to
be in any way representative of real galaxies, since they miss some
key physical processes by construction, namely AGN feedback, metal
evolution and disc instabilities. On the other hand, some of the
excluded physical mechanisms are known to have a strong impact on
predicted galaxy properties: since each model fine-tuning is done on
versions including these additional processes, they could mask or
reduce any difference due to the processes considered here.

\section{Models}
\label{sec:models}

In the following, we briefly review the merger trees used and describe
the main ingredients of the semi-analytic models focusing on the
physical processes considered. We refer the reader to the original
papers for more detailed descriptions; the not-interested reader may
skip directly to sec.\ref{sec:results}.

It is worth stressing, that we make no effort to reduce the
differences between the ``cooling only'' realizations shown in
DL10. This choice allows us to keep our predictions as close as
possible to the original formulation of the three SAMs under
investigation. However, we consider two different sets of predictions
relative to the \oldurham model, by considering both an isothermal and
a cored profile for the hot halo. This choice allows us to compare
predictions of the fiducial \oldurham model with those of a model that
gives results closer to those obtained using the fiducial cooling
models in \morgana and the \munich SAM. Moreover, in order to exclude
differences due to merger times ($t_{\rm mrg}$) assigned to satellite
galaxies, we focus on a ``no-merger'' realization assuming $t_{\rm
  mrg}=\infty$ for all satellites. We also consider ``instantaneous
merger'' realizations ($t_{\rm mrg}=0$), and we will comment on the
predictions of these models whenever appropriate.

We choose the same Chabrier IMF for all models and we switch off metal
production in our model realizations (i.e. primordial composition is
assumed during the entire evolution of our model galaxies). As a
check, we have rerun our realizations allowing metal enrichment: the
results presented in Section~\ref{sec:results} are modified in the
expected direction (e.g., cooling rates are systematically increased),
but our main conclusions remain valid.

\subsection{The Simulations and Merger Trees}
In this work we take advantage of merger trees extracted from two
large, high-resolution cosmological simulations, namely the Millennium
Simulation (MS hereafter, \citealt{Springel05}) and the Millennium-II
Simulation (MSII hereafter, \citealt{BoylanKolchin09}). The MS follows
$N=2160^3$ particles of mass $8.6 \times 10^8 {\rm M_\odot/h}$ within
a comoving box of size $500 {\rm Mpc/h}$ on a side.  The MS-II follows
the evolution of the same number of particles in a volume that is 125
times smaller than for the MS ($100 {\rm Mpc/h}$ on a side), with a
correspondingly smaller particle mass ($6.9 \times 10^6 {\rm
  M_\odot/h}$). For both simulations, the cosmological model adopted
is a $\Lambda$ cold dark matter (CDM) with $\Omega_m=0.25$,
$\Omega_b=0.045$, $h=0.73$, $\Omega_\Lambda=0.75$, $n=1$, and
$\sigma_8=0.9$. The Hubble constant is parametrised as $H_0 = 100 {\rm
  \, h \, km/s/Mpc}$. Group catalogues were constructed using the
standard friend-of-friend (FoF) algorithm, and each group was then
processed using the algorithm {\sc subfind} \citep{Springel01} to
identify self-bound substructures.

We then consider the FoF merger trees constructed as detailed in DL10,
and the same two sets of trees considered there. A first sample (the
``MW-like'' sample) has been constructed by selecting from the MS-II
100 haloes with $log_{10}(M_{200}/M_\odot)$ between $11.5$ and $12.5$
at $z=0$. Here, $M_{200}$ is defined as the mass within a sphere of
density 200 times the critical density of the Universe at the
corresponding redshift. A second sample of 100 haloes was selected
from the MS by taking haloes that have a number density of $10^{-5}
{\rm h^3 Mpc^{-3}}$ at $z\sim2$, and that end up in massive
groups/clusters at $z=0$. The adopted number density has been chosen
to be comparable to that of submillimetre galaxies at $z\sim2$
\citep[see e.g.,]{Chapman04}. We thus refer to this sample as the
``SCUBA-like'' sample.

\subsection{The \munich model}
The stripped-down version of the \munich model used in this work is
built upon the \citet{DeLuciaBlaizot07} implementation and uses
prescriptions for the star formation and feedback that have been
described in detail in \citet{Croton06}. Cold gas is associated only
to the disc component of model galaxies, and both a ``quiescent'' and
a ``starburst'' mode for star formation are considered. Cold gas
surface densities higher than a given threshold $\Sigma_{\rm crit}$
are required for quiescent star formation to take place. This critical
value may be expressed in terms of the galactocentric distance ($R$)
and the virial velocity of the host halo ($V_{\rm vir}$,
\citealt{Kauffmann96}):
\begin{equation}
\Sigma_{\rm crit} = 120 \times \frac{V_{\rm vir} / 200 \, km s^{-1}}{R
  / kpc} M_\odot pc^{-2} \, ,
\end{equation}
\noindent 
which can be translated into a critical mass, assuming the cold gas is
uniformly distributed over a disc with outer radius $r_{\rm
  disc}$. The disc scale length ($r_s$) is computed using results from
the model by \citet{MoMaoWhite98}, and the outer radius of the disc is
assumed to be $r_{\rm disc} = 3\times r_s$. The critical gas mass for
quiescent star formation to take place is:
\begin{equation}
M_{\rm crit} = 3.8 \times 10^9 \frac{V_{\rm vir}}{200 \, km s^{-1}}
\frac{r_{\rm disc}}{10 Kpc} M_\odot \, .
\end{equation}

\noindent
The star formation rate (SFR hereafter) $\varphi$ is then assumed to
occur at the rate:
\begin{equation}
\varphi_{\rm mun} = \alpha_{\rm mun} \frac{M_{\rm
    cold}-M_{\rm crit}}{\tau^{\rm mun}_{\rm dyn,D}} \, ,
\end{equation}

\noindent
where the star formation efficiency is set to $\alpha_{\rm mun}=0.07$,
and $\tau^{\rm mun}_{\rm dyn,D} = r_{\rm disc}/V_{\rm vir}$ is the
disc dynamical time. The adopted modelling leads to episodic star
formation self-regulating to maintain a level close to that
corresponding to the critical surface density.

In addition to the quiescent mode, the model also allows for a
``collisional starburst'' mode of star formation \citep{Somerville01},
triggered by galaxy mergers. The amount of cold gas converted into
stars through this mode depends on the baryonic (gas + stars) mass
ratio of the merging objects. If it is larger than 0.3, the event is
classified as a ``major'' merger, and all cold gas present in the two
merging galaxies is converted into stars. In the case of a minor
merger, the model assumes that only a fraction $f_{\rm brs}$ of all
available cold gas is converted into stars:

\begin{equation}
f_{\rm brs} = e_{\rm brs} \left( \frac{M_1}{M_2} \right)^{a_{\rm brs}} \, ,
\end{equation}

\noindent
where $M_1/M_2$ represents the baryonic mass ratio ($M_2>3M_1$), and
$a_{\rm brs} = 0.7$ and $b_{\rm brs} = 0.56$ have been chosen to
provide a good fit to the numerical simulations of \citet{Cox04}.
Mergers also involve mass transfer between the disc and bulge
components of the remnant galaxy. A detailed description on how galaxy
mergers affect galaxy morphology is presented in \citet{DeLucia11} and
\citet{Fontanot11}. For the purposes of this work we do not
distinguish between bulge and disc components of model galaxies, and
focus on their global properties.

As for stellar feedback, the \munich model links the amount of cold
gas reheated by SNe ($\Delta M^{\rm mun}_{\rm rht}$) in a given time
interval to the mass of stars formed in the same time-step ($\Delta
M_{\rm sf}$):

\begin{equation}
\Delta M^{\rm mun}_{\rm rht} = \epsilon_{\rm rht} \Delta M_{\rm sf} \, ,
\end{equation}
\noindent
where $\epsilon_{\rm rht}=3.5$. The energy released in the same time
interval ($\Delta E_{\rm SN}$) can be written as:

\begin{equation}
\Delta E_{\rm SN} = 0.5 V^2_{\rm SN} \eta_{\rm SN} \Delta M_{\rm sf} \, ,
\end{equation}
\noindent
where $0.5 V_{\rm SN}^2$ represents the mean energy in SNe ejecta per
unit mass of stars formed ($V_{\rm SN}=630 km s^{-1}$ based on
standard SNe theory and a Chabrier IMF), and $\eta_{\rm SN}=0.35$
parametrises its efficiency in reheating the disc cold gas.

Adding the reheated gas to the hot halo without changing its specific
energy leads to the total thermal energy change ($\Delta E_{\rm
  hot}$):

\begin{equation}
\Delta E_{\rm hot} = 0.5 V_{\rm SN}^2 \times \Delta M^{\rm mun}_{\rm rht} \, .
\end{equation}
\noindent
It is then possible to define an excess energy $E_{\rm exc} = \Delta
E_{\rm SN} - \Delta E_{\rm hot}$. If $E_{\rm exc} < 0$, all reheated
gas is confined within the halo, otherwise a fraction $\Delta M^{\rm
  mun}_{\rm eje}$ of hot gas mass ($M_{\rm hot}$) is ejected from the
parent halo through a ``super-wind'':

\begin{equation}
\Delta M^{\rm mun}_{\rm eje} = \frac{E_{\rm exc}}{E_{\rm hot}} M_{\rm hot} =
\left( \eta_{\rm SN} \frac{V^2_{\rm SN}}{V^2_{\rm vir}} -\epsilon_{\rm
  rht} \right) \Delta M_{\rm sf} \, ,
\end{equation}
\noindent
where $E_{\rm hot} = 0.5 V^2_{\rm vir} M_{\rm hot}$ represents the
total thermal energy of the hot gas. The ejected material may be
reincorporated at later times as the parent DMH grows
\citep*{DeLucia04b}:

\begin{equation}
\dot{M}^{\rm mun}_{\rm rei} = \eta^{\rm mun}_{\rm rei} \frac{M_{\rm
    eje}}{\tau_{\rm dyn,H}} \, ,
\end{equation}
\noindent
where $\eta^{\rm mun}_{\rm rei}=0.5$ is a free parameter which
controls the amount of reincorporation per halo dynamical time,
$\tau_{\rm dyn,H} = R_{\rm vir}/V_{\rm vir}$ ($R_{\rm vir}$ being the
virial radius of the parent halo).

\subsection{\andrew}
In the SAM comparison we presented in DL10 we made use of a
stripped-down version of the \citet{Bower06} implementation of the
\oldurham model. There we tested two versions of the \oldurham model,
the difference lying in the assumptions made for the profile of the
hot gas distribution: we defined a ``standard'' model with a
$\beta$-profile (as used by \citealt{Bower06}) and a ``modified''
version using an isothermal profile.

In this work, we take advantage of the new \andrew code
\citep{Benson12} to recreate the stripped-down versions of the
\oldurham models of DL10. \andrew is designed to be highly modular to
facilitate the exploration of different descriptions of key physical
ingredients. We set-up \andrew to run with the same
assumptions\footnote{In the \andrew realization used in this work, all
  other relevant assumptions, including the definition of formation
  times, DM profiles and concentrations, merger time calculation,
  galaxy size calculations, and major merger definitions are treated
  as in the original \oldurham model.}  regarding gas cooling and gas
profiles as in the standard and modified versions of the \oldurham
code used in DL10. In the following, we will refer to these
realizations as \durham (for the $\beta$ or cored profile) and \moddur
(for the isothermal profile). We explicitly check that the ``cooling
only'' stripped down versions of \andrew reproduce with good
approximation the predictions of the ``cooling only'' \oldurham model
presented in DL10. We then include the same treatment for star
formation and stellar feedback as depicted in \citet{Cole00} and
\citet[][see also \citealt{Bower06}]{Benson03}, in order to create an
equivalent of the \oldurham model needed for our analysis. Note that
both the \durham and \moddur models differ significantly from the
default model in the \andrew toolkit.

Galaxy sizes play a key role in this model since they determine
dynamical times and rotation speeds in discs and spheroids. Given the
angular momentum of cooling gas, the radii of galactic discs are
computed by solving for the equilibrium radius at which rotation
supports them against gravity in the combined potential of disc,
spheroid and NFW dark matter halo (including the effects of adiabatic
contraction). When spheroids are formed through major mergers (see
below) their radii are computed by assuming conservation of (internal
plus orbital) energy. Full details are given in \citet{Cole00}.

Star formation is assumed to occur in galactic discs, their cold gas
reservoirs being depleted at a rate determined by the star formation
timescale $\tau_{\rm sf}^{\rm dur}$:

\begin{equation}
\varphi_{\rm dur} = M_{\rm cold} / \tau_{\rm sf}^{\rm dur} \, .
\end{equation}

The star formation timescale is a function of the disc circular
velocity $V^{\rm HR}_{\rm disc}$ taken at the half-mass radius $r^{\rm
  HR}_{\rm disc}$:
\begin{equation}\label{eq:sfdur}
\tau_{\rm sf}^{\rm dur} = \alpha^{-1}_{\rm dur} \tau_{\rm dyn,D}^{\rm
  dur} \left( \frac{V^{\rm HR}_{\rm disc}}{200 km s^{-1}}
\right)^{\beta_{\rm dur}} \, ,
\end{equation}
\noindent
where $\tau_{\rm dyn,D}^{\rm dur} = r^{\rm HR}_{\rm disc}/V^{\rm
  HR}_{\rm disc}$ represents the dynamical time of the disc, while
$\alpha_{\rm dur}=0.0029$ and $\beta_{\rm dur}=-1.5$ are free
parameters.

When two galaxies merge, the merger is deemed to be ``major'' if the
less massive of the galaxies is at least $30\%$ of the mass of the
more massive galaxy. In such cases, the stars of both galaxies are
redistributed into a spheroid. Major mergers always trigger a
starburst. Minor mergers may also trigger a starburst if the gas
fraction in the more massive galaxy exceeds $10\%$. In starbursts, the
gas content of the merging galaxies is placed into the spheroid
component of the merger remnant, where it proceeds to form stars on a
timescale given by eq.~\ref{eq:sfdur} but with $\alpha_{\rm dur}=0.5$,
$\beta_{\rm dur}=0$, and with a dynamical time and velocity computed
for the spheroid, using the method described by \citet{Cole00}.

Stellar feedback is modelled by assuming that the rate at which cold
gas is reheated ($\dot{M}^{\rm dur}_{\rm rht}$) is directly related to
the SFR:

\begin{equation}
\dot{M}^{\rm dur}_{\rm rht} = \left( \frac{V^{\rm HR}_{\rm disc}}{V_{\rm hot}}
\right)^{-\gamma} \varphi_{\rm dur} \, ,
\end{equation}
\noindent
where $V_{\rm hot}=485 km s^{-1}$ and $\gamma=3.2$ are free
parameters. The reheated gas is not instantaneously returned to the
hot phase, but it is stored into a separated reservoir, $M_{\rm
  rsv}$. This material is then added back to the hot phase at a rate
equal to:

\begin{equation}
\dot{M}^{\rm dur}_{\rm rei} = \eta^{\rm dur}_{\rm rei} \frac{M_{\rm
    rsv}}{\tau_{\rm dyn,H}} \, ,
\end{equation}
\noindent
where $\eta^{\rm mun}_{\rm rei}=1.26$ and $\tau_{\rm dyn,H}$ is the
dynamical time of the dark matter halo. It is worth noting that the
reheated material in \andrew behaves as the ejected material in the
\munich model, i.e. as a separated gas phase, which does not take part
in the exchange of energy and mass inside the parent DMH, and which is
slowly reincorporated into the hot phase. For a sake of simplicity, in
the following, we will then refer to this component as ejected
material:

\begin{equation}
\dot{M}^{\rm dur}_{\rm eje} =\dot{M}^{\rm dur}_{\rm rht} \, .
\end{equation}

\subsection{\morgana}\label{morgana}

The MOdel for the Rise of GAlaxies aNd Agns ({\sc morgana}) was first
presented in \citet{Monaco07}. The treatment of star formation and
stellar feedback in this code follows the results of the multiphase
model for the ISM by \citet{Monaco04}.  For the purposes of this work
we consider the combination of parameter values adopted when using a
Chabrier IMF \citep{LoFaro09}.

Discs are treated as ``thin'' systems: SN remnants blow out of the
disc soon after they form and most of the SNe energy is injected into
the halo hot gas, and only a few percent of SNe energy injected into
the ISM. The star formation timescale is of the form:

\begin{equation}
\tau_{\rm sf,D}^{\rm mor} = 9.1 \left( \frac{\Sigma_{\rm cold,D}}{1
  M_\odot pc^{-2}} \right)^{-0.73} \left( \frac{f_{\rm cold,D}}{0.1} 
\right)^{0.45} Gyr \, ,
\end{equation}

\noindent
where $f_{\rm cold,D}$ represents the cold gas fraction (in the disc
component).  Gas re-heated by stellar feedback is ejected from the
disc to the an external reservoir of baryons (i.e. the ``halo''
component) still bound to the host DM halo, at a ``hot wind'' rate
$\dot{M}_{\rm hw}$ assumed to be equal to the star formation rate:

\begin{equation}\label{mor_rht}
\dot{M}_{\rm hw,D} = \varphi_{\rm mor,D} \, .
\end{equation}

\noindent
A fraction $f_{\rm th,D}=0.32$ of SN energy $E_{\rm SN}$ is carried
away with this ejected material, so the contribution $E_{\rm hw}$ of
the hot wind to the thermal energy of the hot halo gas is:

\begin{equation}
\dot{E}_{\rm hw,D} = f_{\rm th,D} E_{\rm SN} 
\frac{\varphi_{\rm mor,D}}{m_{\rm SN}} \, .
\end{equation}

\noindent
where $m_{\rm SN}$ represents the mass of newly formed stars per SN.
A further energy contribution is added by assuming that one type Ia SN
per year explodes each $10^{12}$ M$_\odot$ of stellar mass. This
rather crude implementation of energy from SNe Ia does not influence
much model results.

Cold gas flows into the bulge component either by mergers or by disc
instabilities. Moreover, a fraction of the cooling flow is allowed to
fall directly into the bulge. Star formation in bulges is assumed to
take place in a ``thick'' regime, where SN energy is effectively
trapped within the ISM. In this case, SFR is assumed to follow a
Schmidt-Kennicutt relation, with a timescale:

\begin{equation}
\tau^{\rm mor}_{\rm sf,B} = 4 \times \left( \frac{\Sigma_{\rm
    cold,B}}{M_\odot pc^{-2}} \right)^{-0.4} Gyr \, .
\end{equation}

\noindent
The size of the starburst, necessary to compute the gas surface
density, is estimated as follows. Gas is assumed to have no angular
momentum, and thus to be supported by turbulence generated by SNe.
Under very simple assumptions \citep{LoFaro09}, the velocity
dispersion in this case can be written as:

\begin{equation}\label{eq:sigma}
\sigma_{\rm cold} = \sigma_0 \left( \frac{\tau^{\rm mor}_{\rm
    sf}}{Gyr} \right)^{-1/3} \, ,
\end{equation}

\noindent
where $\sigma_0 = 60$ km s$^{-1}$ is treated as a free parameter.  It
is then assumed that the size of the starburst region is such that
$\sigma_{\rm cold}$ equates the rotation curve of the bulge, assumed
to be flat (see \citealt{LoFaro09} for details).

The rate at which hot gas is ejected from a bulge to the host halo is
assumed to be:

\begin{equation}
\dot{M}_{\rm hw,B} = \left\{ 
\begin{array}{cl}
\frac{\sqrt{V^2_{\rm hot}-V^2_B}}{V_{\rm hot}} \, \varphi_{\rm mor,B} &
\textrm{if $V_B < V_{\rm hot}$} \nonumber \\ 0 & \textrm{if $V_B \ge
  V_{\rm hot}$}
\end{array} \right.
\end{equation}

\noindent
This is done to take into account the ability of the potential well of
a massive bulge to keep hot gas confined; the parameter $V_{\rm hot}$
is chosen to be 300 km/s, corresponding to the typical thermal
velocity of a $\sim10^7$ K hot phase.  The corresponding energy
carried by the hot wind is:

\begin{equation}
\dot{E}_{\rm hw,B} = f_{\rm th,B} E_{\rm SN} \frac{\sqrt{V^2_{\rm
      hot}-V^2_B}}{V_{\rm hot}} \frac{\varphi_{\rm    mor,B}}{m_{\rm SN}} \, ,
\end{equation}

\noindent
where we use $f_{\rm th,B}=0.1$ for the fraction of SN energy carried
away.

A bulge also ejects cold gas into the halo, due to the same SN-driven
turbulence that sets the starburst size (Eq.~\ref{eq:sigma}). This
``cold wind'' from the bulge to the halo component is assumed to occur
at a rate:

\begin{equation}
\dot{M}_{\rm cw,B} = \frac{M_{\rm cold,B} P_{\rm ub} v_{\rm ub}}{R_B} \, ,
\end{equation}

\noindent
where $R_B$ is the half-mass radius of the bulge, while $P_{\rm ub}$
and $v_{\rm ub}$ represent the probability that a cold cloud is
unbound (i.e. it has a velocity larger than the escape velocity of the
bulge $V_B$) and the average velocity of unbound clouds (both
probabilities are computed assuming a Maxwellian distribution of
velocities with rms $\sigma_{\rm cold}$).

The hot and cold halo gas components keep track of both the mass and
the energy received respectively from hot winds from discs and bulges
and from cold winds from bulges. The halo receives winds both from the
central galaxy and from satellites.  Whenever the gas phases of the
halo component are too energetic to be bound to the DM halo, they are
allowed to escape to the IGM as a galaxy ``superwind''. In particular,
if the energy $E_{\rm hot,H}$ of the halo hot gas mass $M_{\rm hot,H}$
overtakes the virial energy $E_{\rm vir}$ by more than a factor
$f_{\rm wind}=2$, a superwind occurs at a rate:

\begin{equation}
\dot{M}^{\rm mor}_{\rm eje,H} = \left( 1-\frac{f_{\rm wind} E_{\rm
    vir}}{E_{\rm hot,H}} \right) \frac{c_s M_{\rm hot,H}}{R_{\rm vir}} \, ,
\end{equation}

\noindent
where $c_s$ represents the sound velocity in the halo.  A similar
formula determines the ejection of cold superwinds:

\begin{equation}
\dot{M}^{\rm mor}_{\rm eje,C} = \left( 1-\frac{f_{\rm wind} V^2_{\rm
    disp}}{\sigma^2_H} \right) \frac{\sigma_H M_{\rm cold,H}}{R_{\rm
    vir}} \, .
\end{equation}

A fraction ($f_{\rm back}=0.5$) of the mass ejected by the DM halo is
later re-incorporated (i.e. added to infalling IGM), when the parent
DM halo reaches an escape velocity larger than the (thermal or
kinetic) velocity the gas had when it was ejected.

\subsection{Comparison between the models}

All three models considered relate the SFR in the disc component to
the amount of cold gas there available, but make different assumptions
for the timescale of conversion. In the \munich model, the star
formation proceeds on a timescale directly proportional to the
dynamical time of the disc, while in \oldurlike models the timescale
of star formation is rescaled with some power of the circular velocity
of the disc. Finally, \morgana assumes a star formation timescale
consistent with the Schmidt law. The \munich model explicitly accounts
for a critical mass threshold for star formation, while \morgana and
\oldurlike models do not.

Besides a quiescent mode of star formation, both the \munich and
\durham models assume an enhanced star formation regime occurring
during galaxy mergers, with a fraction of the total cold gas available
being turned into stars in a very short timescale (that of the model
integration). In {\sc morgana}, cold gas is associated also to the
bulge component. Because of the complete loss of angular momentum, gas
in bulges is concentrated to very small sizes. Then the higher gas
surface densities force the Schmidt-Kennicutt relation to give much
higher SFRs and shorter star formation timescales than those
corresponding to star formation in discs.  In our reference runs, we
explicitly exclude merger events and the enhanced SFR by setting
$t_{\rm mrg}=\infty$. We will discuss the impact of the ``starburst''
mode on the predicted star formation histories by considering the
predictions of the instantaneous-merger runs in section~\ref{sec:sfh}.

In \morgana the amount of disc gas reheated via stellar feedback
($M_{\rm rht}$) equals the SFR ($\varphi$), while in the \munich model
these two quantities are proportional, via the parameter
$\epsilon_{\rm rht}>1$. This implies that the amount of reheated gas
per unit star formation is always larger in the \munich model than in
\morgana. A slightly different choice has been made in the \andrew
models, where the amount of reheated gas remains proportional to
$\varphi$, but is additionally assumed to scale as a power of the disc
velocity.

An important quantity in the balance of the baryonic content of each
DMH is given by the amount of baryons ejected. In both \oldurlike and
\munich models, the fraction of ejected mass is directly proportional
to the mass of stars formed in the same time interval, while in
\morgana the dependency of the ejection rate on stellar feedback is
mediated via the estimate of the thermal (kinetic) energy of the hot
(cold) gas halo phase. All models assume that this ejected material is
reincorporated into the halo at later times, with different
assumptions for the reincorporation rates. In \oldurlike and \munich
models, this is modelled as a continuous process following the DMH
growth, while in \morgana only half of the material connected to each
ejection event is re-acquired instantaneously when the DMH has grown
enough to overcome its escape velocity.

\oldurlike and \munich models employ a very similar scheme for
modeling the ejected component. In both cases some of the feedback
reheated material is excluded from the mass/energy flows between the
cold and hot gas phases: but while this is only a fraction in the
\munich model (the remaining being instantaneously added to the hot
phase), all reheated material is considered in the \oldurlike
models. The same reincorporation scheme is assumed, but with a faster
rate in the \oldurlike models than in the \munich model.

In the following, we choose to avoid the complications arising form
disc instabilities\footnote{In particular, disc instabilities have no
  direct effect on the SFR predicted by the \munich model, since only
  enough stellar mass is removed from the disc to the bulge to restore
  stability. In the \oldurham model, at each instability event, the
  whole disc is destroyed and all its baryons are given to the
  spheroidal component, with any cold gas present assumed to undergo a
  starburst. In \morgana a different choice has been made, by moving a
  well defined fraction of disc stars and cold gas to the bulge, where
  it forms stars on a $\tau_{\rm sf,B}^{\rm mor}$ timescale.} by
switching them off.

\section{Results}\label{sec:results}
\begin{figure*}
  \centerline{ \includegraphics[width=16cm]{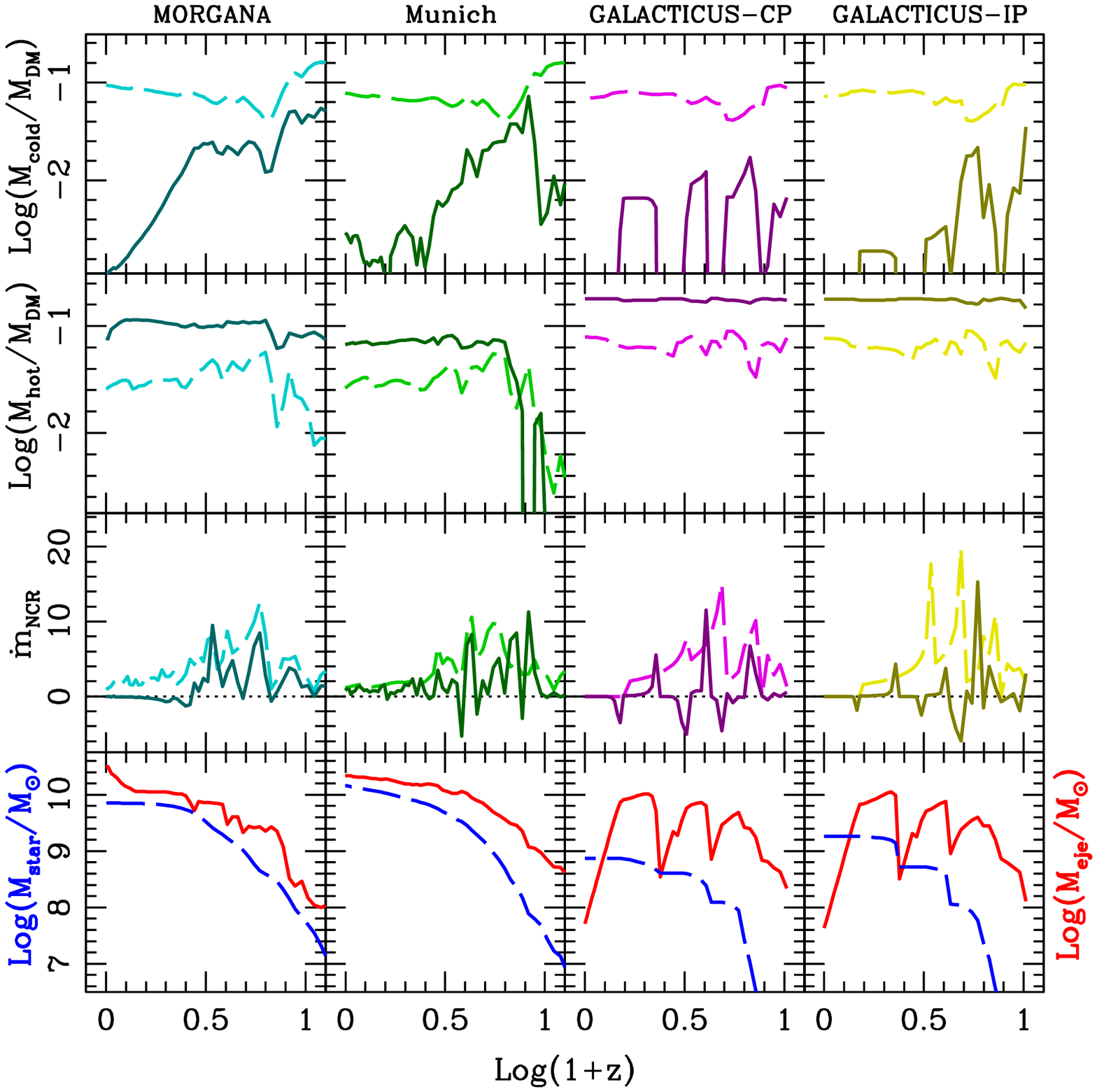} }
  \caption{Redshift evolution of the baryonic content for a MW-like
    representative DMH (with quiet mass accretion history; fig.1 right
    panels in \citealt{DeLucia10}). {\it Upper row}: cold gas
    fractions associated with the central galaxy; {\it second row}:
    hot gas fractions; {\it third row}: net cooling rates. Blue, red,
    yellow and green lines refer to \morgana, \durham, \moddur and the
    \munich model. Dark solid lines refer to the models considered in
    this work, while dashed lines refer to the ``cooling only'' models
    \citep{DeLucia10}. {\it Lower row}: stellar masses (blue dashed
    lines) and ejected masses (red solid lines). In all models we
    assume $t_{\rm mrg}=\infty$ (see text for more
    details).}\label{fig:mii66}
\end{figure*}
\begin{figure*}
  \centerline{ \includegraphics[width=16cm]{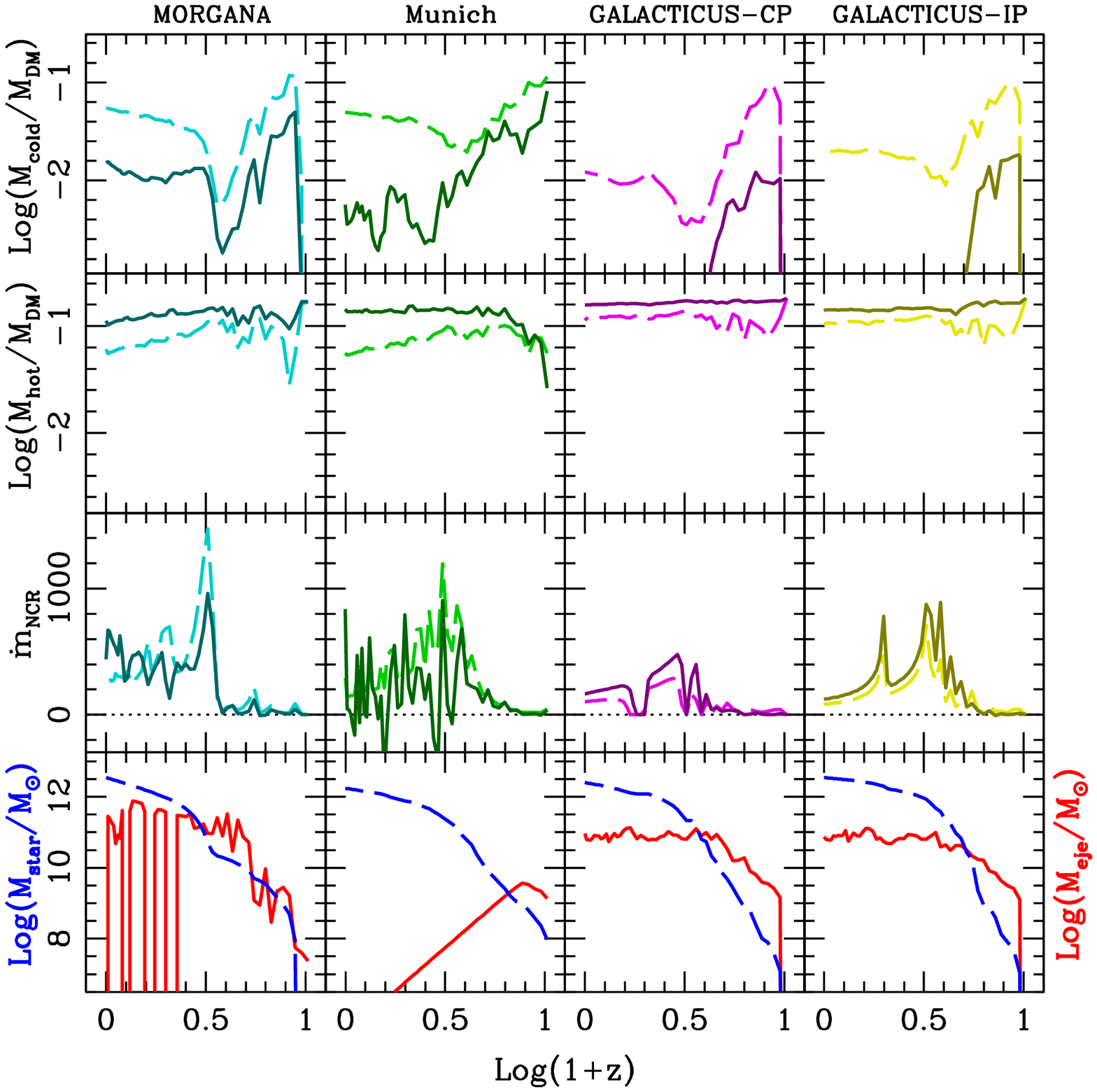} }
  \caption{Redshift evolution of the baryonic content for a SCUBA-like
    representative DMH (with quiet mass accretion history; fig.4 right
    panels in \citealt{DeLucia10}). Symbols, line styles, and colours
    have the same meaning as in
    fig.~\ref{fig:mii66}.}\label{fig:scb06}
\end{figure*}

\subsection{Test cases}\label{sec:repre}

We start by considering the evolution of the different baryonic
components in a few test cases. For consistency with DL10, we focus on
the same 4 representative DMHs (their fig.~1 and~4). These have been
chosen among the 100 MW-like and the 100 scuba-like (two for each
sample) as representative of a quiet mass accretion histories and of a
large number of merger events. In the following, whenever we refer to
``central galaxy'' for our DMHs, we will refer to the central galaxy
of the main progenitor at the corresponding redshift. All other
galaxies will be defined as ``satellites''. For the sake of
simplicity, in this section we will show results only for the 2 haloes
with quiet mass accretion histories. The other two haloes give
consistent results, and are shown for completeness in
appendix~\ref{appA}.

Figures~\ref{fig:mii66} and~\ref{fig:scb06} show the redshift
evolution of the different baryonic components in the haloes
considered: from top to bottom, we show the cold gas fraction
associated with the central galaxy, the hot gas fraction in the halo,
the {\it net cooling rate} ($\dot{m}_{\rm NCR}$; see below), the
stellar mass of the central galaxy and the mass ejected from the
DMH. In the three upper panels, we compare the predictions of the
models including star formation and feedback (darker colours - solid
lines) with previous results for the ``only cooling'' realizations
considered in DL10 (lighter colours - long dashed lines).

We define $\dot{m}_{\rm NCR}$ as the net rate at which cold gas and
stars are deposited into the galaxy:

\begin{displaymath}
\begin{array}{cc}
\dot{m}_{\rm NCR} = & \frac{M_{\rm cold}(t_2)+M_{\rm
    star}(t_2)}{t_2-t_1} - \frac{M_{\rm cold}(t_1)+M_{\rm
    star}(t_1)}{t_2-t_1} \, ,
\end{array}
\end{displaymath}
\noindent
where $t_1$ and $t_2$ are the cosmic epochs corresponding to two
consecutive snapshots. The net cooling rate differs from the intrinsic
cooling rate, since it takes into account both the effect of feedback
in removing part of the cold gas from the system and the effect of
star formation in locking some material in long lived stars. For the
stripped-down SAM versions considered in DL10 these two quantities
coincide by construction.

The inclusion of feedback from star formation has the net effect of
reducing the amount of cold gas available in MW-haloes
(figure~\ref{fig:mii66}, this holds also for instantaneous merger
realizations) and increasing the hot gas fraction with respect to the
``cooling only'' runs. In the MW haloes, this increase is particularly
relevant in the \oldurlike models, that predicts higher hot gas
fraction with respect to the other two models. This shows that the
feedback scheme implemented in the \oldurlike models is the most
efficient (among those considered here) in reheating the cold gas in
the haloes at these mass scales. The cold gas fraction associated with
the central galaxy is characterized by a marked decline at lower
redshift, and predictions from the three models considered are more
different than in the ``cooling only'' realizations. In our models
there are two competing effects able to deplete the cold gas
reservoir: the formation of long lived stars and cold gas removal by
stellar feedback. In order to disentangle these two effects we
consider the evolution of the overall cold gas plus stars and still
find a decrease with respect to the predictions of DL10 for the cold
gas component. These findings show that the inclusion of a strong
stellar feedback is the main driver of the different evolution of the
baryonic components for the models considered in this study.

The combined effect of the decreased cold gas and increased hot
component is shown in the third row. In general, the evolution of
$\dot{m}_{\rm NCR}$ is not smooth and characterized by epochs of {\it
  negative} contribution, i.e. time intervals dominated by outflows
able to reduce the cold gas content of the central object. For the
MW-like haloes, the evolution of $\dot{m}_{\rm NCR}$ show large
deviations from the ``cooling-only'' configuration, with extended
redshift ranges characterized by low or negative rates. This is
particularly evident for the \oldurlike realizations, while the
\munich model gives predictions that are closest to the cooling rates
obtained in the ``cooling only'' runs.

In the lower panels of fig.~\ref{fig:mii66}, we show the redshift
evolution of the baryonic mass ejected from the DMH (red solid line)
and the stellar mass associated with the central galaxy (blue dashed
line). The first quantity is crucial to keep track of the total amount
of baryons in each main progenitor, and gives important hints on the
impact of stellar feedback. At this mass scale, the ejected component
dominates over the stellar mass deposited in the central galaxy for
MW-like haloes, \citep[see also][]{DeLucia04b}. We stress that the
ejected mass in the \oldurlike realizations is comparable to the
results for the other two models for the MW-like haloes, confirming
that the differences in the cold and hot gas fractions are mainly due
to the different feedback efficiency and not to an enhanced fraction
of material excluded from the mass/energy flows. Indeed, the similar
amounts of ejected mass in \oldurlike and \munich models in the whole
redshift range, despite $\eta^{\rm dur}_{\rm rei} \gg \eta^{\rm
  mun}_{\rm rei}$, confirms that the stellar feedback scheme adopted
in \oldurlike models removes a larger fraction of gas from the
baryonic budget of these haloes with respect to the \munich scheme.

In the same panel we also show the evolution of the central galaxy
stellar mass: at this mass scale and with the inclusion of only star
formation and stellar feedback, model results are very different. The
\munich model predicts the largest stellar masses at all redshifts,
\morgana gives slightly lower values for the stellar mass, and both
\oldurlike runs predict stellar masses almost one order of magnitude
smaller than the other two models. The \moddur model predicts slightly
larger stellar masses with respect to the \oldurlike models. These
results for the \oldurlike models are consistent with our analysis of
the net cooling rate, and are due to a feedback-driven starvation of
the cold gas reservoir of the central galaxy. We checked that these
conclusions also hold for the redshift evolution of the mean stellar
mass in the central galaxy (averaged over the 100 realizations in each
DMH sample).


Similar conclusions are reached when considering the representative
SCUBA halo (figure~\ref{fig:scb06}), but with some significant
differences. For these haloes, both the increase of hot gas fraction
and the decrease of the cold gas fraction with respect to the
``cooling only'' versions are less marked, and the prediction of
different models are somewhat closer. This is due to the fact that
stellar feedback is more efficient in affecting the thermal state of
the gas and galaxy evolution on a MW-like scale than in more massive
haloes. Consistently, in SCUBA-like haloes $\dot{m}_{\rm NCR}$ follows
much more closely the evolution of cooling rates in DL10, and the
outflow dominated periods are less frequent or completely absent. As
already noticed in DL10, net cooling rates can takes up values of
several hundreds of M$_\odot$ yr$^{-1}$ with the exception of the
\durham model, where spikes are much less pronounced.

At these mass scales, the ejected component does not dominate over the
stellar mass for the SCUBA-like sample \citep{DeLucia04b} at $z<2$,
and the different feedback models predict a rather different
evolution: in the \oldurlike runs the amount of material in the
ejected component stays constant below $z\sim 3$, a more noisy
evolution is seen for \morgana, while in the \munich model the ejected
mass decreases rapidly and is negligible for the baryonic budget of
the haloes at present. The differences in the predictions of the three
models are due to the different interplay between the ejected and
reincorporated fraction, and again shows that in \oldurlike models the
fraction of ejected material is larger than in the \munich model,
leading to an almost constant ejected mass, despite the faster
reincorporation rate.

Finally, the stellar masses predicted by the three SAMs for the
SCUBA-like sample are much closer than in the MW-like sample, with
only \moddur predicting slightly higher stellar masses. We interpret
also this result as an effect of a less efficient stellar feedback in
SCUBA-like haloes for \oldurlike models: in these models gas cooling
is always dominating over cold gas removal (third row) and more
material for star formation is available. Therefore, in this case the
marked decrease of the cold gas fraction associated with the central
galaxy is the effect of a more efficient star formation.

\subsection{Star Formation Histories}\label{sec:sfh}
\begin{figure*}
  \centerline{ \includegraphics[width=16cm]{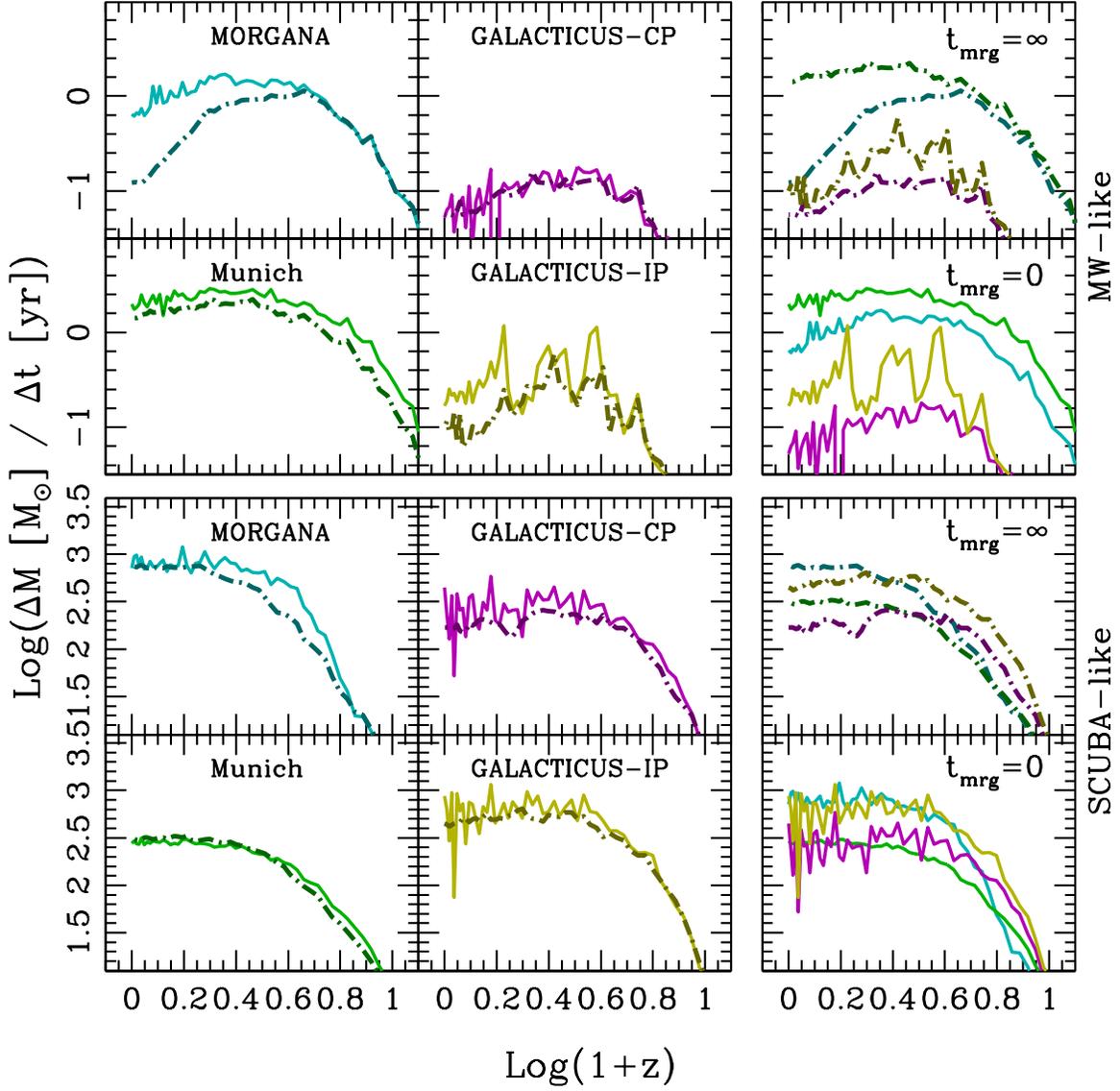} }
  \caption{Average evolution of central galaxy stellar mass as a
    function of redshift (over 100 DMHs - see text for more
    details). Blue, red, yellow and green lines refer to \morgana,
    \durham, \moddur and the \munich model. Dark dot-dashed lines
    refer to the $t_{\rm mrg}=\infty$ realizations, while light solid
    lines correspond to the instantaneous merger runs. Upper panels
    refer to the MW-like sample, while lower panels to the SCUBA-like
    sample. Left panels compare the no merger and the instantaneous
    merger runs for each SAM separately, while right panels compare
    the average prediction of different SAMs in the same
    configuration.}\label{fig:mean_SFR}
\end{figure*}

We now turn our attention to the stellar mass assembly of the central
galaxy as predicted by our SAMs. In figure~\ref{fig:mean_SFR}, we
consider the evolution, averaged over the whole sample of 100 halos,
of the stellar mass in the central galaxies between two contiguous
snapshots: in our reference models, with no galaxy mergers, this
corresponds to the mean SFR in the object between the two snapshots
(dark lines). In order to quantify the relative contribution of the
``starburst'' mode to central galaxy mass assembly, we also consider
the predictions of the instantaneous merger run ($t_{\rm mrg}=0$) and
proceed as follows. We first compute the mean stellar mass variation
for central galaxies in the runs with instantaneous mergers: these
quantities include both the SFRs in central galaxies and the
contribution from satellite mergers. In the no-merger runs, we then
consider the mean stellar mass content in satellites already accreted
by the main progenitor at each redshift. We then subtract this from
the mean stellar mass variations for central galaxies in instantaneous
merger runs.  This quantity (shown as light lines in the figure) does
not strictly correspond to the SFR in the central galaxy in the
instantaneous merger runs, since enhanced SFR episodes in satellites
are not properly subtracted. Clearly, the instantaneous merger runs
correspond to an extreme case, since in a realistic SAM run only a
fraction of satellites is allowed to merge onto their central
galaxy. Nonetheless, the comparison of dark and light line provides a
conservative upper limit to the overall starburst mode contribution to
the assembly of the central object.

In each panel of figure~\ref{fig:mean_SFR}, we consider the mean
variation of stellar mass in the two different sets of runs, averaged
over the 100 MW-like (upper panels) and SCUBA-like haloes (lower
panels). In the left panels, we directly compare the no merger runs
with the instantaneous merger runs modified as described above for
each SAM, to show the contribution of the starburst mode
in each configuration. In the right panels, we compare the
different predictions of our SAMs in the same runs.

Let us focus first on the MW-like sample. As far as the no merger runs
are considered, \morgana and the \munich model predictions are very
close to each other at early times, and start diverging at lower
redshifts, with \morgana predicting less star formation than the
\munich model. At the same mass scales, the \andrew realizations show
lower SFRs at all redshifts. With respect to the \munich model, the
difference is about one order of magnitude, while at lower redshifts
the difference with \morgana predictions is reduced. The \moddur
predicts systematically higher SFRs at all redshift than the \durham
run. These results are consistent with our findings in
sec.~\ref{sec:repre}, and are directly related to the differences in
net cooling rates among the SAMs. We compare these results with the
analogous prediction for the instantaneous merger runs, thus showing
the maximum expected contribution from the starburst mode of star
formation. In general the overall growth rate of the central galaxy is
enhanced in all SAMs. The largest contributions are seen at low
redshift for \morgana (at $z<3$, reaching a factor of 8 at $z\sim0$)
and for \moddur (a factor of 4 at $z<1$), while the \munich model
shows a smaller (at most a factor of 2) increase at all redshift. The
smallest modifications are seen for \durham (only a few percent).

We then focus the SCUBA-like sample. For the no merger runs, the
predictions of the different models are generally closer than for the
MW-like sample. At these mass scales, the predictions of the different
\andrew realizations are more similar to those from the \munich model
and \morgana, and they show higher SFRs at earlier times ($z \gtrsim
2$). Again, \moddur realizations predicts systematically higher SFRs
with respect to \durham realizations. Also for this sample, the
instantaneous merger runs predict an enhancement of the stellar mass
formed, but in this case the variations are well below a factor of two
in all cases, with the maximum deviation seen for \morgana at
intermediate ($z\sim3$) redshifts (it is worth recalling that the two
examples of figures~\ref{fig:scb06} and \ref{fig:scb48} show spiky
deposition rates for three out of four models; these differences are
not visible in the SFR when it is averaged over 100 halos). These
results shows that, at these mass scales, the impact of merger-driven
starburst is limited. This is in line with the phenomenological
estimate of \cite{Sargent12} of an 8-14 per cent contribution of
mergers to the total SFR.

\subsection{Stellar and gas content at different redshifts}

\begin{figure*}
  \centerline{ 
    \includegraphics[width=8cm]{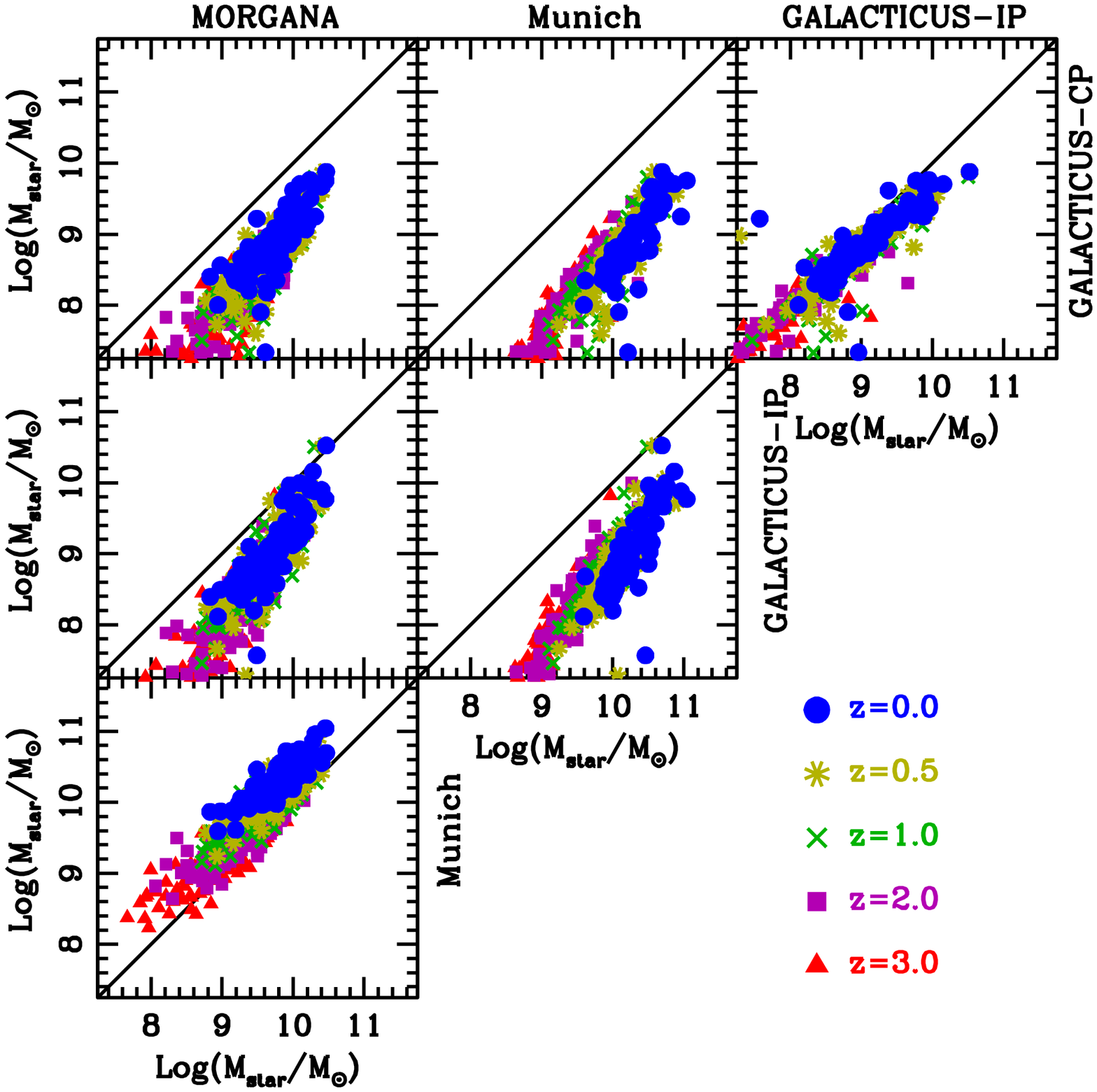} 
    \includegraphics[width=8cm]{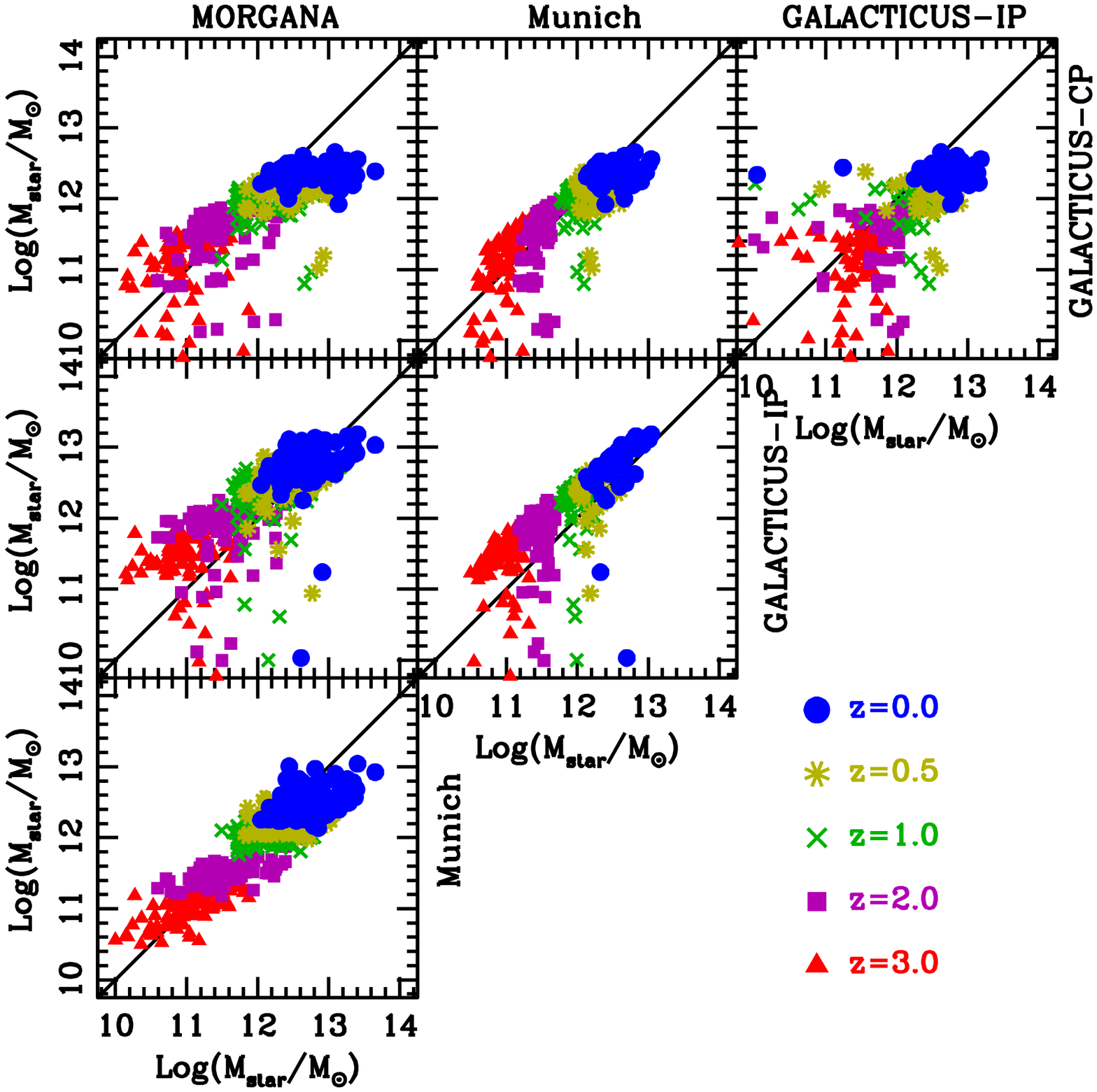} 
  }
  \caption{Comparison of central galaxy stellar masses for the 100
    MW-like ({\it left panel}) and SCUBA-like({\it right panel})
    haloes. In each panel, blue circles, yellow asterisks, green
    crosses, purple squares and red triangles refer to model
    predictions at $z=[0; 0.5; 1; 2; 3]$
    respectively.}\label{fig:distr_star}
\end{figure*}

In order to get more insight into the mass assembly process predicted
by our models, we compare in fig.~\ref{fig:distr_star} the model
stellar masses for the central galaxy at different redshift on an
object-by-object basis. Left and right panels refer to the MW-like and
SCUBA-like haloes, respectively. Each pair of models show some degree
of correlation in the predicted stellar masses, and this confirms the
overall consistency of the models. For the MW-like haloes, the stellar
content predicted by the \oldurlike realizations is offset low with
respect to predictions from the other models. Moreover, the overall
slope of the relation looks steeper than the one-to-one correlation,
with smaller galaxies deviating more from the one-to-one
relation. This shows that the difference is due to the mass depencence
of the efficiency of feedback in regulating star formation. For the
SCUBA-like haloes, the stellar masses predicted from the three models
considered are very close. \durham is the only model deviating
considerably from the one-to-one relation, particularly at low
redshift, reflecting the fact that these galaxies grow more slowly in
the \durham realizations than in \morgana or the \munich model. We
also consider the corresponding predictions for the ``instantaneous
merger'' run (not shown in the figure): we find that all central
galaxies in SCUBA-like haloes at all redshift lie along the one-to-one
line, while our conclusions are qualitatively unchanged in MW-like
haloes. For these runs, the scatter in the correlations is reduced in
both samples at all redshift: we thus conclude that the presence of
mergers helps the predictions of the models to converge to a common
value, and this effect gets stronger with increasing number of
mergers.
\begin{figure*}
  \centerline{ 
    \includegraphics[width=8cm]{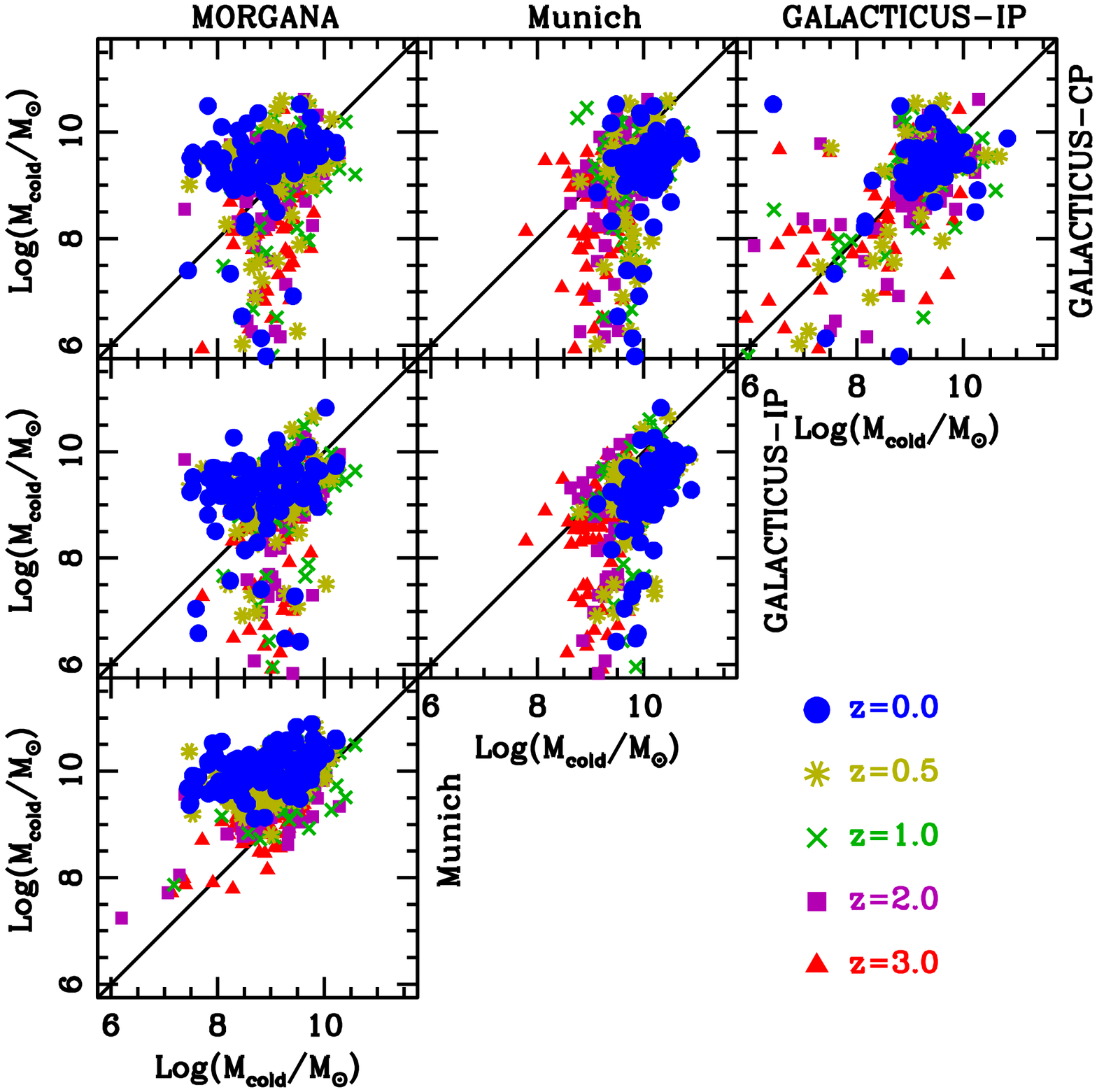} 
    \includegraphics[width=8cm]{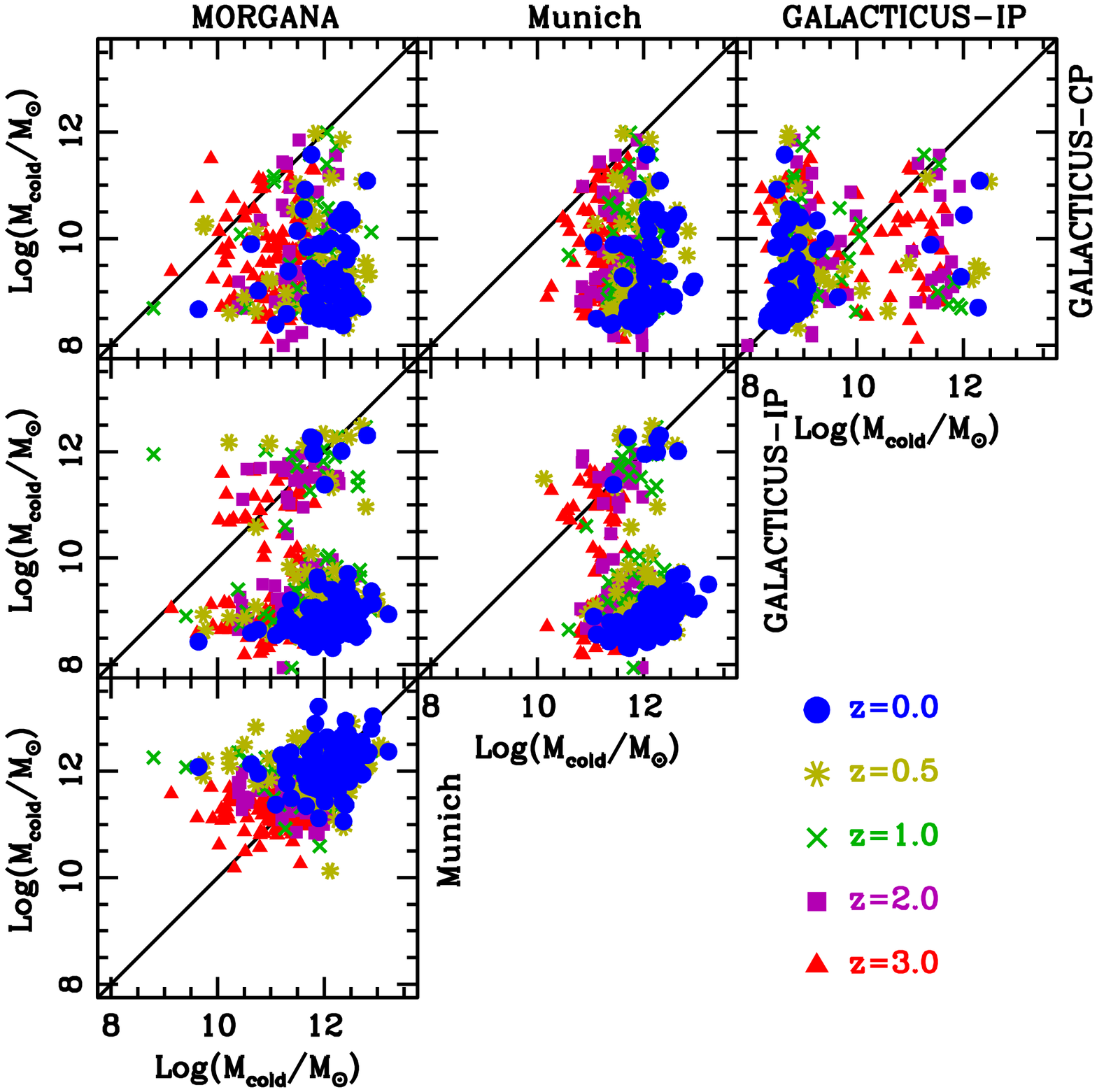} 
  }
  \caption{Same as fig.~\ref{fig:distr_star} for cold gas masses
    associated with the central galaxy in the 100 MW-like ({\it left
      panel}) and SCUBA-like({\it right panel}) haloes. Symbols and
    colours are the same as in
    fig.~\ref{fig:distr_star}.}\label{fig:distr_gas}
\end{figure*}

We also consider the corresponding cold gas content associated with
the central galaxy on an object-by-object basis
(fig.~\ref{fig:distr_gas}). The differences between the SAMs are even
more evident for this physical quantity in both samples. \morgana and
the \munich model show the strongest correlation both for the MW-like
and SCUBA-like haloes, while the \oldurlike realizations show no
correlation at all with the predictions of the other two models. It is
worth stressing that in SCUBA-like haloes \oldurlike realizations
predict cold gas amounts $1-2$ orders of magnitude lower than the
corresponding \morgana and \munich predictions. The results are
similar between \durham and \moddur, implying that this depletion is
not only related to the smaller cooling flows associated with the
cored-profile, but that stellar feedback is also playing an important
role in removing cold gas content from the haloes. However, it is not
possible to indicate stellar feedback as the only responsible for the
smaller stellar masses obtained in MW-like haloes. Indeed, the cold
gas content of these haloes in the \oldurlike runs may be even larger
than the corresponding \morgana and \munich predictions. Therefore, it
is the interplay between star formation and feedback that is
responsible for the different predictions.

\section{Discussion \& Conclusions}
\label{sec:final}

We compared predictions of three independently developed semi-analytic
models of galaxy formation, focusing our analysis on their assumed
modelling for the physical processes involving star formation and
stellar feedback. Following the same approach as in
\citet[DL10]{DeLucia10} we define ``stripped down'' SAM versions
including cooling, star formation, feedback from supernovae (SNe) and
simplified prescriptions for galaxy merging, and we run them on the
same samples of DMH merger trees, extracted from the Millennium and
Millennium-II Simulations. 
 
The choice of ``stripped-down'' versions of the SAMs has the advantage
of avoiding complications due to other processes like disc
instabilities, metal enrichment and AGN feedback. In our
``stripped-down'' versions we either assume $t_{\rm mrg}=\infty$ or
$t_{\rm mrg}=0$, in order to remove the additional degeneracies due to
different definitions of merging times (see DL10). Our aim is to
discuss the influence of specific model ingredients on the physical
properties of model galaxies. In the following, we summarise and
discuss our main findings.

\begin{itemize}
\item{{\it Supernovae feedback:} As expected, switching on star
  formation and stellar feedback has important consequences on the
  different gas phases in DMHs, with respect to the ``cooling only''
  SAM versions: the amount of cold gas available is reduced, while the
  hot gas fraction is increased.  While we find an encouraging level
  of consistency between the models, the specific star formation and
  SN feedback prescriptions in each model induce differences in the
  stellar and gaseous content of galaxies. This is in line with what
  is found in numerical simulations of MW-like galaxies performed with
  different codes \citep{Scannapieco12}.  In particular, we find that
  the scheme adopted by the \oldurlike models provides larger hot gas
  fractions with respect to the other two models and a rapid depletion
  of the cold gas fraction in galaxies. The amount of gas that is in
  an `ejected' component (i.e. temporarily not associated with the
  halo/galaxy) is similar in the three models considered, though its
  redshift evolution can differ significantly.}

\item{{\it Stellar content:} If we consider the average SFRs in the
  MW-like sample, \morgana and the \munich model predictions show a
  good level of agreement at $z\gtrsim2$: predictions from the two
  models deviate at later times, which has only a limited effect on
  the predicted $z=0$ stellar masses. For this sample, the most
  striking differences are between predictions of \morgana (or the
  \munich model) and those from the \oldurlike realizations. In fact,
  the star formation and feedback schemes implemented following
  \citet{Bower06} predict significantly lower SFR levels at all
  redshifts. As a consequence, the \oldurlike models predicts
  systematically lower stellar masses for the corresponding central
  objects with respect to both \morgana and the \munich model. The
  discrepancy is a weak function of stellar mass itself, being smaller
  for more massive galaxies (within a given DMH sample). For
  SCUBA-like haloes, the differences between the predicted SFRs are
  smaller than for MW-like haloes, and the predicted stellar masses
  are much closer at all redshifts, the biggest difference being a
  more rapid assembly of stellar mass in the central object at earlier
  times in the \oldurlike runs.}

\item{{\it Quiescent and starburst modes of star formation:} We
  analysed the impact of the ``starburst'' mode of star formation
  (usually associated with galaxy mergers) by comparing runs obtained
  with infinite and vanishing galaxy merging times; this approach
  allows us to give an upper limit on the contribution of the
  starburst mode.  In most cases we found an increase of the average
  mass assembly rate of the central galaxy not larger than a factor of
  2, with the largest contributions found in the \morgana model. This
  shows that the contribution of the starburst mode of star formation
  in the overall SFR budget is limited, in line with, e.g.,
  \citet{Sargent12}. We also find that the inclusion of merger driven
  starbursts decrease the scatter in the predicted stellar masses at
  given redshift between haloes in the same DMH sample.}
\end{itemize}

These results extend and deepen our conclusion presented in DL10 to
include the comparison of different approaches to the modeling of star
formation and feedback in SAMs. Despite the general coherent picture
for the impact of the energy injection connected to stellar feedback
on the distribution of baryons into the different gas phases, each
feedback model is characterized by its unique pattern in tracing the
redshift evolution of the different baryonic components (cold gas, hot
gas, ejected gas) and this information is fundamental to understand
the building up of galaxy properties in a cosmological context.

\section*{Acknowledgements}
FF acknowledges financial support from the Klaus Tschira Foundation
and the Deutsce Forschungsgemeinschaft through Transregio 33, ``The
Dark Universe''. GDL acknowledges financial support from the European
Research Council under the European Community's Seventh Framework
Programme (FP7/2007- 2013)/ERC grant agreement n. 202781. MBK
acknowledges support from the Southern California Center for Galaxy
Evolution, a multi-campus research program funded by the University of
California Office of Research. Some of the calculations were carried
out on the ``Magny'' cluster of the Heidelberger Institute f\"ur
Theoretische Studien. The Millennium and Millennium-II Simulation data
bases and the web application providing online access to them were
constructed as part of the activities of the German Astrophysical
Virtual Observatory.  We are grateful to Gerard Lemson for setting up
an internal data base that greatly facilitated the exchange of data
and information needed to carry out this project.

\bibliographystyle{mn2e}
\bibliography{fontanot}

\appendix
\section{Predictions for representative haloes with merger-dominated mass
  accretion histories.}\label{appA} For consistency with DL10, we show
in this appendix the two DMHs chosen among the 100 scuba-like and the
100 MW-like as representatives of haloes with a significant number of
merger events along their assembly history (compare with results in
sec.~\ref{sec:repre}).
\begin{figure*}
  \centerline{ \includegraphics[width=16cm]{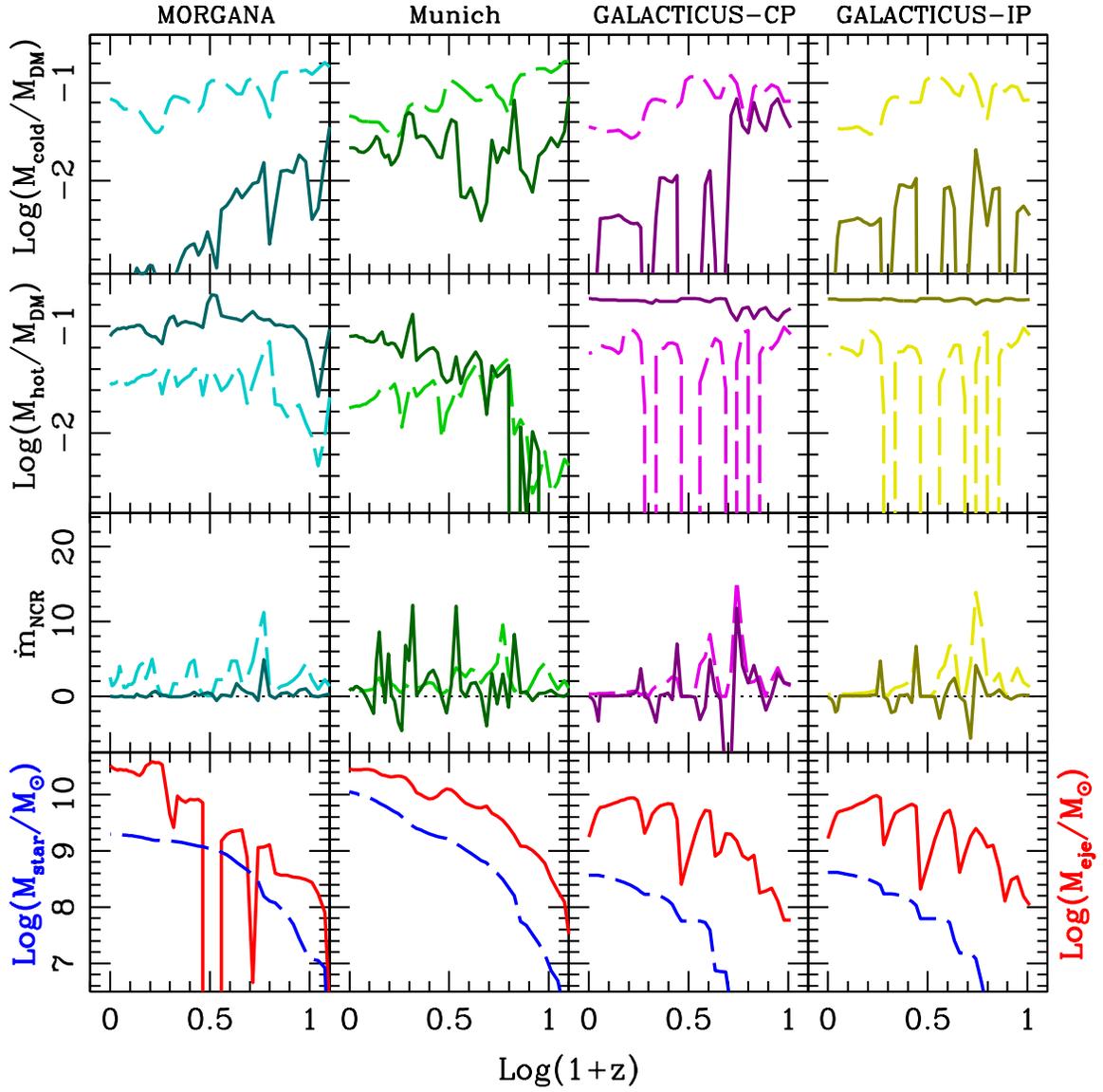} }
  \caption{Redshift evolution of the baryonic content of a MW-like
    representative DMH (with merger-dominated mass accretion history;
    fig.1 left panels in \citealt{DeLucia10}). Symbols, line styles,
    and colours are as in fig.~\ref{fig:mii66}.}\label{fig:mii57}
\end{figure*}
\begin{figure*}
  \centerline{ \includegraphics[width=16cm]{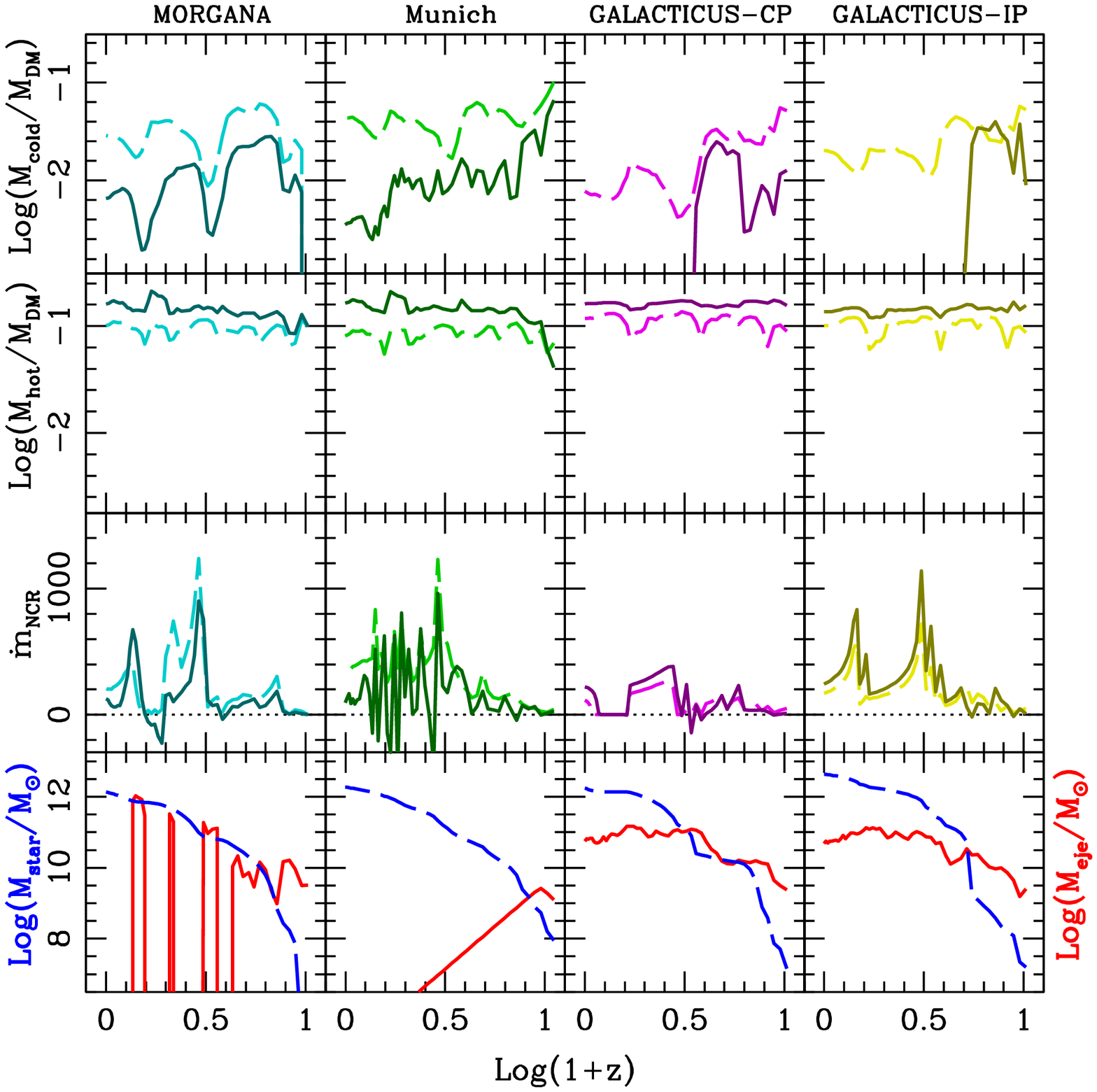} }
  \caption{Redshift evolution of the baryonic content of a SCUBA-like
    representative DMH (with merger-dominated mass accretion history;
    fig.4 left panels in \citealt{DeLucia10}). Symbols, line styles,
    and colours are as in fig.~\ref{fig:mii66}.}\label{fig:scb48}
\end{figure*}

\label{lastpage}

\end{document}